\documentclass[aps,prb,10pt,twocolumn,floatfix,superscriptaddress,showpacs]{revtex4-1}
\usepackage{graphicx}
\usepackage{bm}
\usepackage{amssymb}
\usepackage[usenames]{color}
\usepackage{epsfig}
\usepackage{subfigure}

\usepackage{soul}
\setulcolor{red}
\setstcolor{red}
\sethlcolor{yellow}

\newcommand{\be}{\begin{equation}}
\newcommand{\ee}{\end{equation}}
\def \be{\begin{equation}}
\def \ee{\end{equation}}
\def \ba{\begin{array}}
\def \ea{\end{array}}
\def \bea{\begin{eqnarray}}
\def \eea{\end{eqnarray}}

\begin{document}

\title{Manipulating Majorana fermions using supercurrents}

\author{Alessandro Romito}
\affiliation{\mbox{Dahlem Center for Complex Quantum Systems and Fachbereich Physik, Freie Universit\"at Berlin, 14195 Berlin, Germany}}
\author{Jason Alicea}
\affiliation{Department of Physics and Astronomy, University of California, Irvine, CA 92697, USA}
\author{Gil Refael}
\affiliation{Department of Physics, California Institute of Technology, Pasadena, CA 91125, USA}
\author{Felix von Oppen}
\affiliation{\mbox{Dahlem Center for Complex Quantum Systems and Fachbereich Physik, Freie Universit\"at Berlin, 14195 Berlin, Germany}}

\begin{abstract}
Topological insulator edges and spin-orbit-coupled quantum wires in proximity to $s$-wave superconductors can be tuned through a topological quantum phase transition by a Zeeman field. Here we show that a supercurrent flowing in the $s$-wave superconductor also drives such a transition. We propose to use this mechanism to generate and manipulate Majorana fermions that localize at domain walls between topological and nontopological regions of an edge or wire.  In quantum wires, this method carries the added benefit that a supercurrent reduces the critical Zeeman field at which the topological phase appears.      
\end{abstract}


\maketitle

{\em Introduction.---}Emergent Majorana fermions in a condensed matter setting are currently attracting much attention.\cite{wilczek09,Franz,Stern,Service,Hughes} Zero-energy Majorana fermions comprise the simplest non-Abelian anyon and promise fascinating applications to topological quantum information processing.\cite{kitaev03,freedman98} Recently, the set of candidate systems supporting Majorana fermions has greatly expanded beyond quantum Hall systems \cite{review,read00} with the realization that several materials can be driven into a topological superconducting phase when placed in proximity to a conventional $s$-wave superconductor.  This was initially understood for topological insulators,\cite{fu08,fu09} followed by 2D $s$-wave superfluids with Rashba spin-orbit interaction, \cite{sato09} spin-orbit-coupled quantum wells \cite{sau10,alicea10} and nanowires,\cite{lutchyn10,oreg10} half-metals,\cite{duckheim11,lee09,chung11} and 3D topological insulator nanoribbons.\cite{MarcelProposal}

Nanowire proposals are attractive as they involve widely available materials and provide detailed recipes for manipulating the Majorana fermions\cite{alicea11}---a prerequisite for verifying their non-Abelian statistics and performing topological quantum information processing. While initial proposals examined simple mean-field models of clean wires proximate to a superconductor, more recent work indicates that the induced topological phase persists in the presence of moderate interactions \cite{gangadharaiah11,sela11,stoudenmire11} or disorder,\cite{brouwer11a,brouwer11b,potter10,potter11,stanescu11} and considered setups for probing the Majorana bound states.\cite{fu09,leijnse11,wimmer11,jiang11,driss11} Experimental challenges nevertheless remain: realizing the topological phase requires control over the wire's global electron density and the application of significant Zeeman fields without destroying superconductivity.  Furthermore, manipulating Majorana fermions by \emph{locally} controlling the electron density using gate electrodes\cite{oreg10,alicea11,hassler11} is nontrivial due to strong screening by the superconductor.

\begin{figure}[b]
\includegraphics[width=8.5cm]{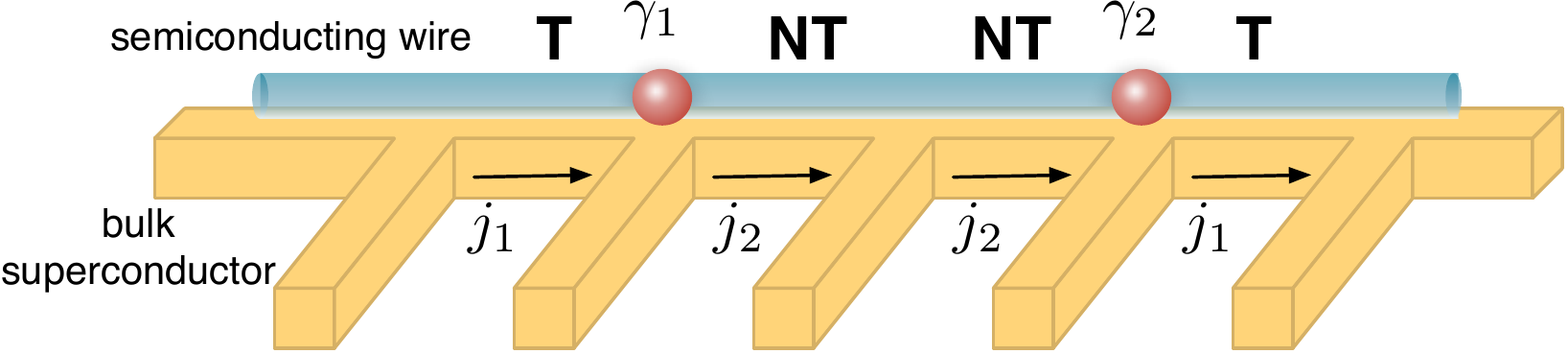}
\caption{Schematic device for manipulation of Majorana fermions by supercurrents. Different sections of the proximity-inducing $s$-wave superconductor carry different currents, putting the quantum wire into an alternation of topological (T) and nontopological (NT) phases. Changing the supercurrent in the segments allows for manipulation of the Majorana fermions $\gamma$ associated with the domain walls.}
\label{fig:device}
\end{figure}

Here we show that the latter two challenges can be greatly alleviated by applying supercurrents in the bulk superconductor. These supercurrents cause a spatial gradient of the phase of the proximity-induced pair potential in the wire, which drives a transition between the nontopological and topological superconducting phases. Remarkably, the supercurrent also allows one to access the topological phase at weaker Zeeman fields.
Spatially varying the phase gradient along the wire moreover generates {\em controllable} Majorana-carrying domain walls between nontopological and topological regions. Switching the supercurrents (and hence the phase gradient) along sections of the quantum wire transports these Majorana fermions preserving the gap, obviating the need for local gating. Schematically, this can be achieved by the device shown in Fig.\ \ref{fig:device}.  Our scheme, in fact, applies equally well to the edge of a two-dimensional topological insulator with proximity-induced superconductivity.\cite{fu09,seradjeh11}  We start by elucidating the physics in this setting, since it is somewhat simpler to analyze, and then turn to the quantum wire case.

\begin{figure}
\includegraphics[width=6.5cm]{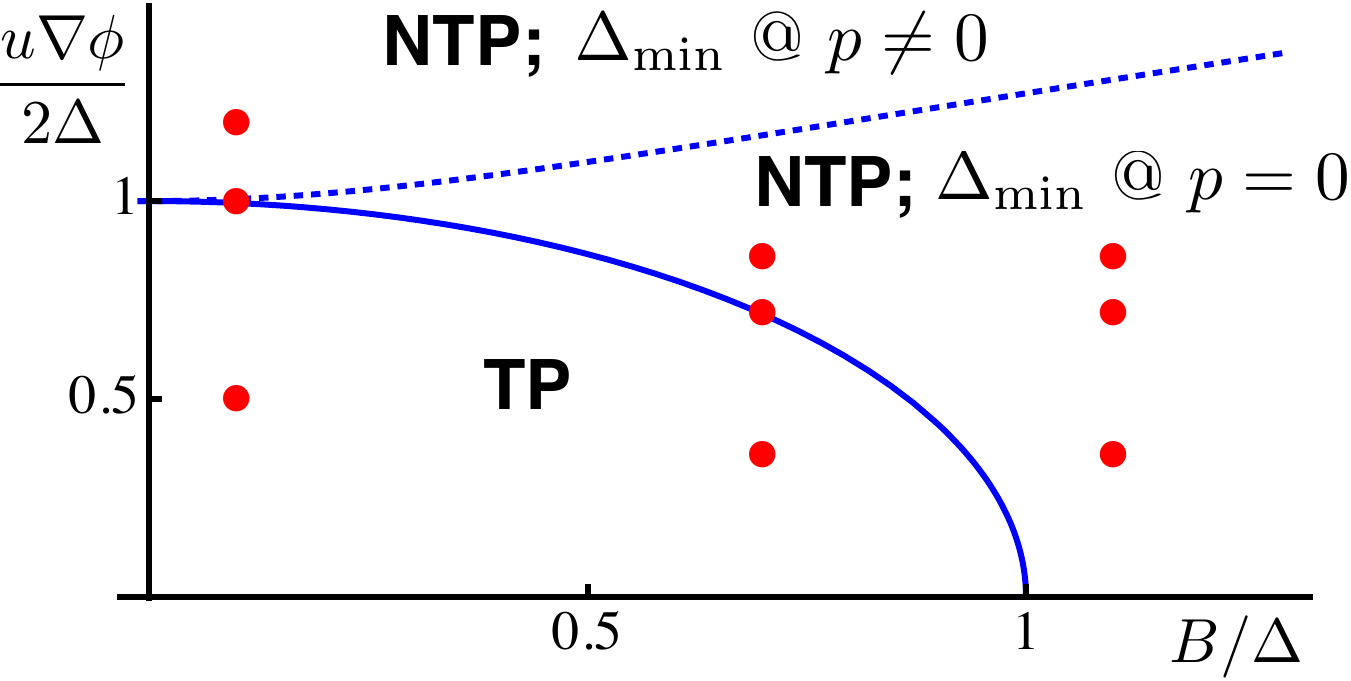}
\caption{ Phase diagram for the topological insulator case with $\mu=0$. Increasing the Zeeman field $B$ or the superconducting phase gradient $\nabla\phi$ induces a phase transition from the topological phase (TP) into a nontopological phase (NTP). The latter has two regimes, separated by the dotted line, depending on whether the minimal excitation energy $\Delta_{\rm min}$ is located at zero or nonzero $p$. The red dots indicate representative points at which the spectrum is illustrated in Fig.\ \ref{fig:top-ins}.}
\label{fig:phases}
\end{figure}

{\em Topological insulators.---}Our analysis begins from the Bogoliubov-de Gennes equation describing a topological insulator edge proximate to an $s$-wave superconductor and subjected to a magnetic field:\cite{fu08,fu09}
\begin{equation}
{\cal H} = (up\sigma_x -\mu) \tau_z -B \sigma_z + \Delta e^{i\phi(x)}\tau_+  + \Delta e^{-i\phi(x)}\tau_- .
\label{Htopins}
\end{equation} 
Here, $p$ is the momentum along the edge, $u$ measures the edge-state velocity, $\mu$ is the chemical potential, $\Delta$ and $\phi(x)$ denote the magnitude and (position-dependent) phase of the proximity-induced pair potential, and $B\geq 0$ is the Zeeman field. The Pauli matrices $\sigma_i$ and $\tau_i$ respectively act in spin and particle-hole space. Equation (\ref{Htopins}) is written in a Nambu basis with spinors of the form $\psi = [u_\uparrow, u_\downarrow, v_\downarrow, -v_\uparrow]^T$.  

In the absence of a phase gradient, say $\phi=0$, the spectrum of Eq.\ (\ref{Htopins}) is easily derived by repeatedly squaring $\cal H$, exploiting the fact that the Bogoliubov spectrum is symmetric about zero energy.\cite{fu08} Focusing for simplicity on $\mu=0$ until specified otherwise, one finds energies $E_\pm(p) = \{(up)^2 + |\Delta \pm B|^2\}^{1/2}$.  With $B<\Delta$, $\cal{H}$ describes a topological superconducting state\cite{fu09} similar to that of Kitaev's model for a 1D $p$-wave superconductor.\cite{1DwiresKitaev}  The gap closes when $B = \Delta$, signifying a topological quantum phase transition into a trivial superconducting state, and reopens at larger $B$; see Fig.\ \ref{fig:phases}. 

Consider now Eq.\ (\ref{Htopins}) with a nonuniform phase $\phi(x)$. Rather than studying the Hamiltonian (\ref{Htopins}) directly, it is advantageous to gauge away $\phi(x)$ from the pairing term using the unitary transformation 
\begin{equation}
 {\cal U} = \exp\{i \phi(x) \tau_z/2 \}.
\label{gauge}
\end{equation} 
Note that this gauge transformation multiplies the electron and hole components of the Bogoliubov-de Gennes spinor by opposite phase factors. The  transformed Hamiltonian ${\cal H}^\prime = {\cal UHU}^\dagger$ becomes 
\begin{equation}
  {\cal H}^\prime =  \left[u\left(p-\frac{\nabla \phi}{2} \tau_z\right)\sigma_x -\mu\right] \tau_z -B \sigma_z + \Delta \tau_x  ,
\label{H:gauge}
\end{equation} 
which depends only on the gradient of $\phi$, making gauge invariance explicit. 

\begin{figure}[t]
\includegraphics[width=8.5cm]{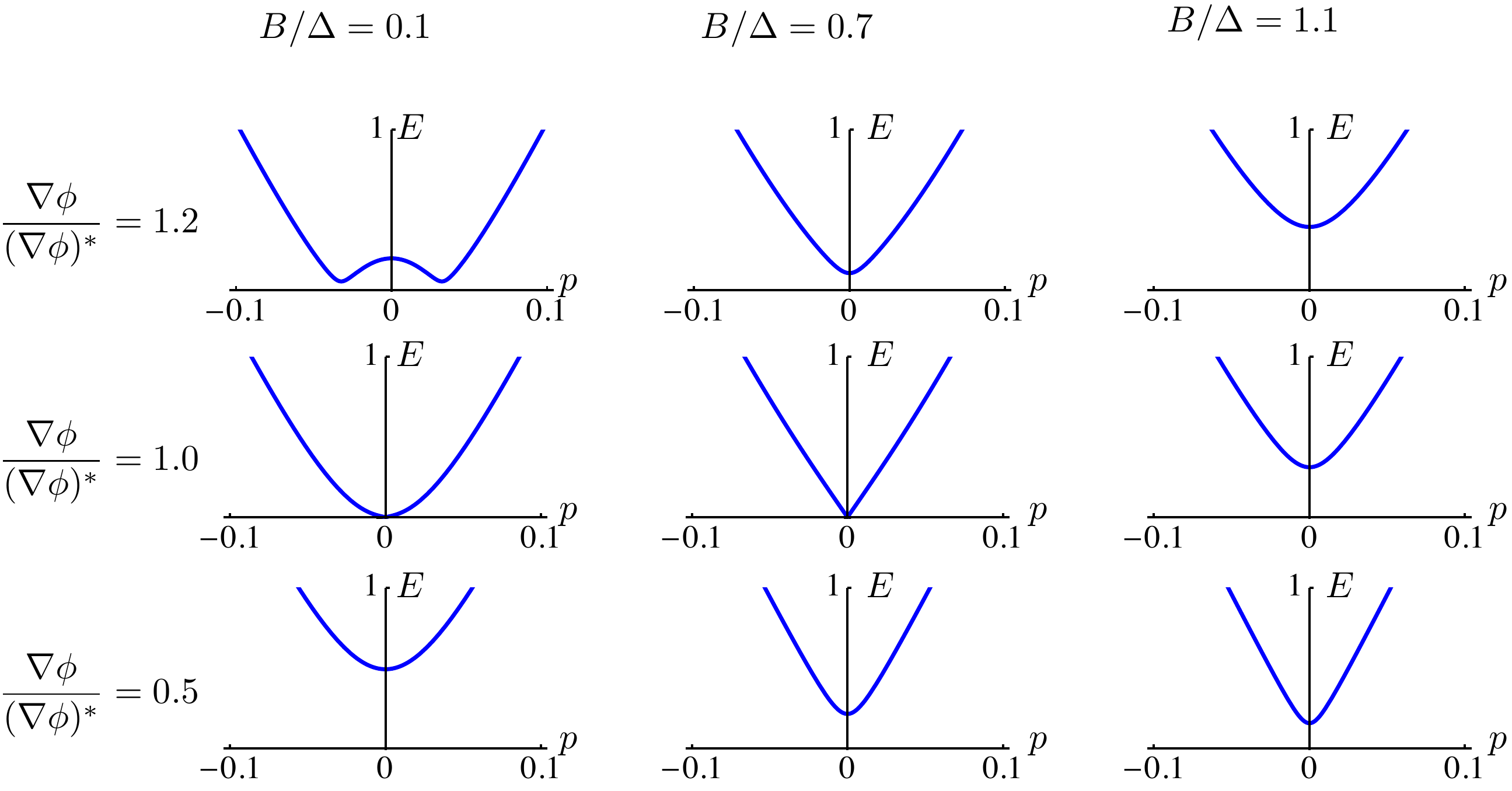}
\caption{Bogoliubov de Gennes spectrum of the topological insulator at $\mu=0$ for various Zeeman fields and superconducting phase gradients, as indicated in the figure. The corresponding points are also marked in the phase diagram of Fig.\ \ref{fig:phases}. By symmetry, only the positive-energy dispersion is shown. $E$ and $p$ are measured in units of $\epsilon_{\textrm{so}}$ and $2m u$, respectively.
}
\label{fig:top-ins}
\end{figure}

For uniform phase gradients $\nabla \phi$ along the wire, the Hamiltonian ${\cal H}^\prime$ in Eq.\ (\ref{H:gauge}) is readily diagonalized; Fig.\ \ref{fig:phases} summarizes the resulting phase diagram. Representative spectra, corresponding to the points in the phase diagram marked in Fig.\ \ref{fig:phases}, appear in Fig.\ \ref{fig:top-ins}. When the system resides in the nontopological phase ($B > \Delta$) one finds an enhancement of the gap as the phase gradient increases from zero. Conversely, when the system begins in the topological phase ($B < \Delta$), the gap initially decreases when applying a nonzero phase gradient. Eventually the gap closes at the critical phase gradient
\begin{equation}
(\nabla \phi)^* = (2\Delta/u)[1-(B/\Delta)^2]^{1/2},
\label{phase_boundary_top_ins}
\end{equation}
which signals the transition into a nontopological phase as indicated by the full line in Fig.\ \ref{fig:phases}.  Note that $(\nabla \phi)^*$ is of the order of the inverse of the proximity-induced (zero-$B$) coherence length of the edge state, $\xi = u/\Delta$.  Increasing the phase gradient beyond $(\nabla\phi)^*$ reopens the gap for any nonzero $B$. (At zero $B$, there is a transition into a gapless phase.) This shows that a gradient of the phase of the gap function, originating from supercurrents in the bulk $s$-wave superconductor, can indeed induce a transition between the topological and nontopological superconducting phase.  One can understand this result intuitively by observing from Eq.\ (\ref{H:gauge}) that
the phase gradient behaves exactly as a magnetic field oriented along $x$, which effectively shifts the electrons' momentum (in contrast to the Zeeman field $B$ which opens a gap).  When $\nabla \phi = 0$ and $\Delta > B$, electrons with opposite momenta are resonant and can easily form Cooper pairs, driving the system to the topological phase with a $\Delta$-dominated gap.  A nonzero phase gradient, however, breaks the resonance between $p$ and $-p$ states and suppresses Cooper pairing.  If this shift is too large, the Zeeman field $B$ dominates the gap and the edge forms a trivial phase.  

When increasing the phase gradient even further in the nontopological phase, there is another characteristic line where the minimal gap in the excitation spectrum is no longer located at $p=0$, but rather at a finite $p$. This crossover line is indicated by the dotted line in Fig.\ \ref{fig:phases}.  

{\em Quantum wire.---}We now turn to the phase diagram for a quantum wire, proximity coupled to a supercurrent-carrying $s$-wave superconductor. Just as for the topological insulator edge, this phase diagram is obtained from the quantum wire Hamiltonian\cite{lutchyn10,oreg10} 
\begin{equation}
{\cal H} = (p^2/2m + up\sigma_x -\mu) \tau_z -B \sigma_z + \Delta e^{i\phi(x)}\tau_+  + \Delta e^{-i\phi(x)}\tau_- 
\label{Hwire}
\end{equation} 
by performing the gauge transformation (\ref{gauge}). In this Hamiltonian, $m$ denotes the effective mass and $u$ the Rashba spin-orbit-coupling strength.   When $\phi(x)$ is uniform, the quantum wire forms a nontopological superconducting phase for $B<\Delta$, undergoes a topological quantum phase transition at $B=\Delta$, and enters a topological superconducting state at larger Zeeman fields.\cite{oreg10,lutchyn10}  

\begin{figure}
\includegraphics[width=6.5cm]{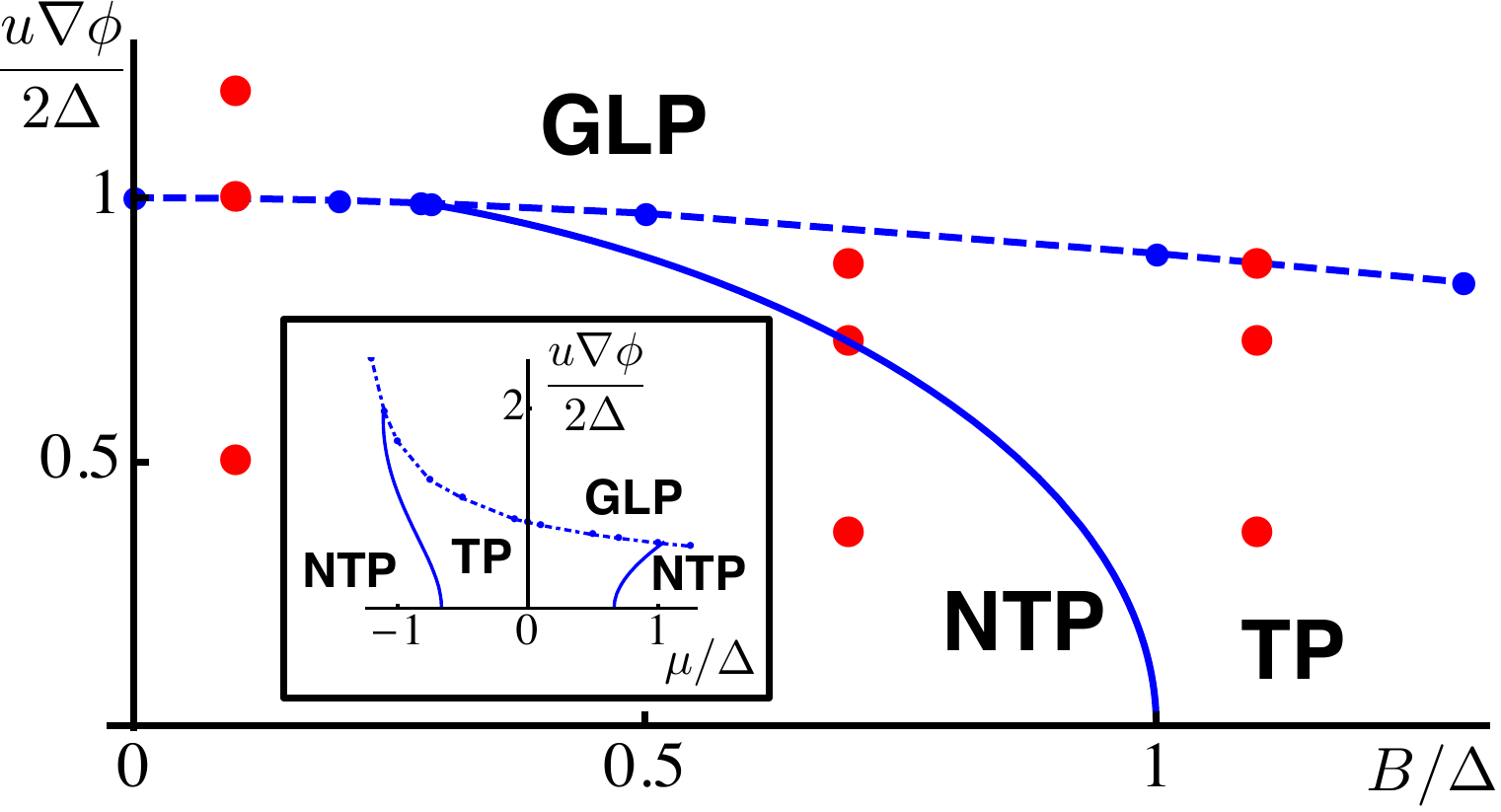}
\caption{Phase diagram of a quantum wire proximity coupled to a supercurrent-carrying $s$-wave superconductor for  $\Delta=\epsilon_{\rm so}$ in the Zeeman field--phase gradient plane at $\mu=0$ (main panel) and the chemical potential--phase gradient plane at $B/\Delta=1.2$ (inset). In both panels the full (dashed) line indicates where the gap vanishes at zero momentum (near $p=\pm p_F$), leading to gapped topological (TP) and nontopological (NTP) phases as well as a gapless phase (GLP). The red dots indicate representative points at which the spectrum is illustrated in Fig.\ \ref{fig:PW2}.}
\label{fig:PDW}
\end{figure}

\begin{figure}
\includegraphics[width=8.5cm]{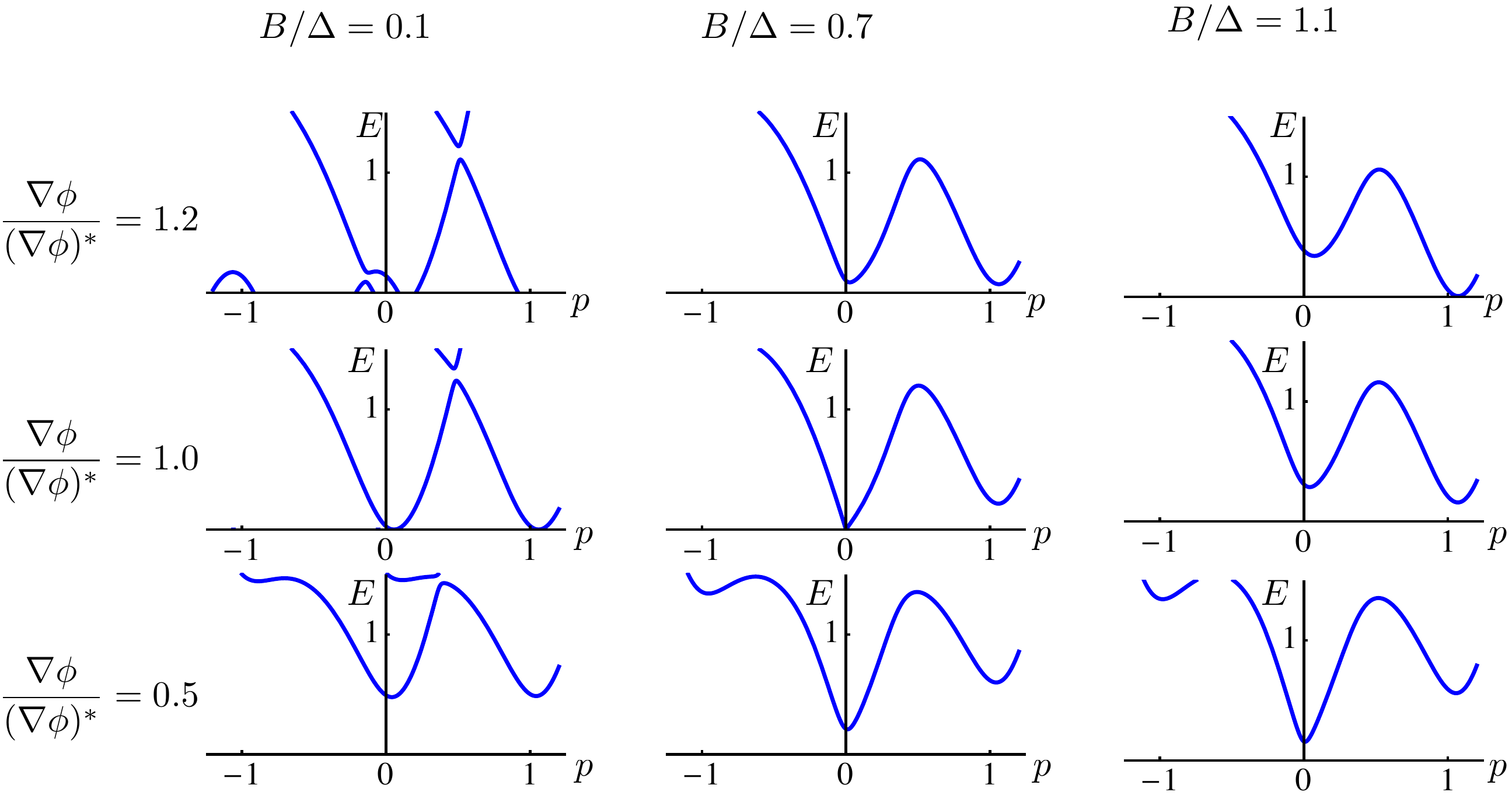}
\caption{Bogoliubov de Gennes spectrumof the quantum wire in proximity to a supercurrent-carrying $s$-wave superconductor, focusing on positive energies due to the symmetry $(p, E) \rightarrow (-p,-E)$. The chemical potential has been set to $\mu=0$; othe parameters are specified above and indicated by red dots in Fig.\ \ref{fig:PDW}. $E$ and $p$ are measured in units of $\epsilon_{\textrm{so}}$ and $2m u$, respectively.}
\label{fig:PW2}
\end{figure}

To understand the resulting phases when a uniform phase gradient is present, we first focus on the excitation spectra for $\mu=0$ in the vicinity of the Fermi points (for $\Delta=B=0$) at $p=0$ and $\pm p_F=\pm2mu$.  Figure \ref{fig:PDW} shows the resulting phase diagram in the Zeeman-field--phase gradient plane. The full line indicates where the gap in the excitation spectrum closes at $p = 0$ while remaining finite on {\em both} sides of the line. The dashed line meanwhile indicates where the gap closes near the Fermi momentum $p_F$. In this case, the gap is finite only on the low phase gradient side of the line, while it remains closed on the high phase gradient side. Figure \ref{fig:PW2} shows representative excitation spectra for the parameters indicated by red dots in the phase diagram of Fig.\ \ref{fig:PDW}(a). These figures show that for $B$ below a critical value $B^*$ a uniform phase gradient drives the wire from the nontopological phase directly into a gapless state.  More interestingly, at $B^* < B < \Delta$ we find that the quantum wire enters the topological phase at intermediate phase gradients before reaching the gapless state at even larger phase gradients. Finally, when $B>\Delta$ the quantum wire forms a topological phase all the way up to a critical phase gradient where it becomes gapless. 

Quantitatively, we find for $\mu=0$ that the phase boundary between the gapped topological and nontopological phases (solid line in Fig.\ \ref{fig:PDW}) is implicitly given by
\begin{equation}
 \left( \frac{u (\nabla\phi)^*}{4 \epsilon_{\textrm{so}}} \right) ^4-4\left( \frac{u (\nabla\phi)^*}{4 \epsilon_{\textrm{so}}} \right)^2= \frac{B^2-\Delta^2}{\epsilon_{\textrm{so}}^2}
\end{equation} 
where $\epsilon_{\rm so} = mu^2/2$. Note that for $|B^2-\Delta^2|\ll \epsilon_{\rm so}^2$, this yields the phase boundary $u (\nabla \phi)^*=2 (\Delta^2-B^2)^{1/2}$ which is independent of $\epsilon_{\textrm{so}}$. For $\Delta\ll\epsilon_{\rm so}$, the junction of the full and the dashed lines in Fig.\ \ref{fig:PDW} can be readily accessed analytically and occurs at $B^*=\Delta^2/4\epsilon_{\rm so}$. 

The inset in Fig.\ \ref{fig:PDW} illustrates the effect of a phase gradient at a nonzero chemical potential for $B>\Delta$. 
At $\nabla \phi=0$ the wire is topological for $|\mu| <\sqrt{B^2-\Delta^2}$. Remarkably, a finite (but not too large) phase gradient extends the topological region to larger $|\mu|$, making the existence of Majorana bound states less sensitive to the tuning of the chemical potential. 

As for the topological insulator edge, these phase diagrams indicate that a phase gradient induces a topological phase transition over a wide region of Zeeman fields, allowing one to realize Majorana-carrying domain walls by spatially varying supercurrents. Moreover, these results show that in the presence of a phase gradient, the Majorana carrying topological phase can be induced in the quantum wire at {\em weaker} Zeeman fields. Both of these facts may significantly simplify the experimental realization of Majorana fermions in quantum wires. 

\begin{figure}
\includegraphics[width=6.5cm]{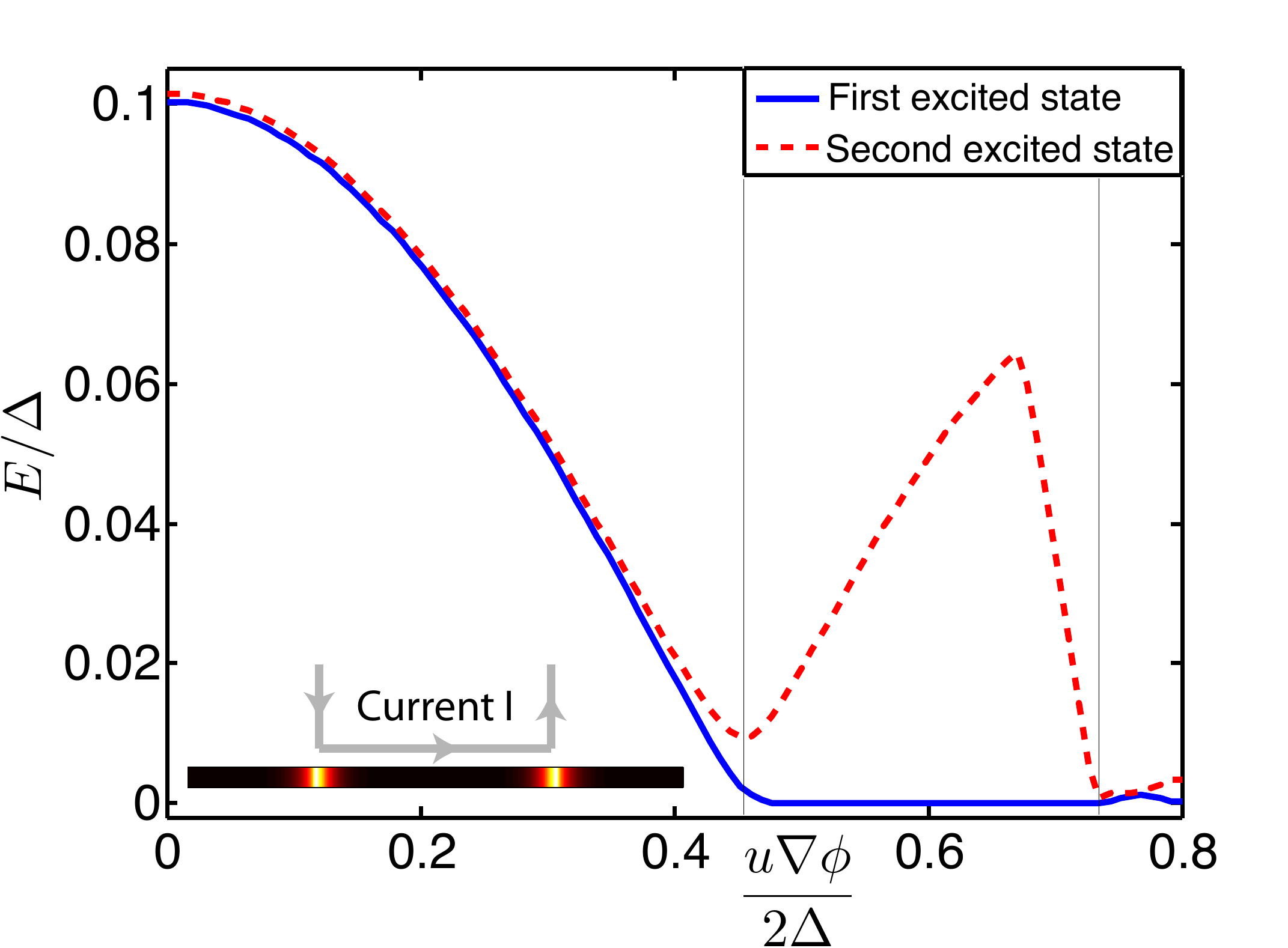}
\caption{Lowest two excitation energies of a 1000-site quantum wire vs.\ the phase twist generated by a supercurrent through the entire wire for $\mu=0$, $B/\Delta=0.9$, $\Delta/\epsilon_{\textrm{so}}=2$.  At zero twist, the wire is in the nontopological phase with $B<\Delta$. At larger twists, the wire enters the gapped topological phase where the lowest excitation energy drops to zero---due to the formation of Majorana modes---before reaching the gapless phase at even higher phase gradients, in line with the phase diagram in Fig.\ \ref{fig:PDW}.  Inset (lower left): Probability distribution for the near-zero mode obtained when current flows only through the central half of the wire, creating a trivial--topological--trivial domain structure with two Majorana modes.}
\label{fig:First}
\end{figure}

To further illustrate the supercurrent-driven transition into a topological phase, we computed the lowest two excitation energies of a finite-length wire subjected to a uniform phase gradient across the entire system using a lattice model that recovers Eq.\ (\ref{Hwire}) in the low-density limit.\cite{stoudenmire11} Figure \ref{fig:First} shows the results for a 1000-site wire with $\mu = 0$, $B/\Delta = 0.9$, and $\Delta/\epsilon_{\textrm{so}} = 2$.  Since $B< \Delta$ here, one finds that both excitation energies remain finite up to a critical phase gradient, indicating that the system forms a nontopological phase. Above the critical phase gradient, the lowest excitation energy drops essentially to zero due the formation of localized Majorana end-states associated with the entry into the topological phase, while the second energy remains finite reflecting the wire's bulk gap.  Beyond a second critical phase gradient, the bulk gap closes and the system enters a gapless phase, in agreement with the phase diagram in Fig.\ \ref{fig:PDW}.  The lower left of Fig.\ \ref{fig:First} displays the probability distribution for the near-zero mode generated by a phase gradient $u\nabla\phi/(2\Delta) = 0.6$ induced only over the central half of the same 1000-site wire.  The phase gradient creates two domain walls between the trivial outer ends and topological inner region, each of which clearly binds a localized Majorana mode as expected.  

{\em Estimates.---}The proposed scheme to manipulate Majorana fermions requires one to establish a sufficient phase gradient to drive the quantum wire (or topological insulator edge) between the topological and nontopological phase without hitting the critical current density of the $s$-wave superconductor. This critical current density corresponds to approximately one $2\pi$ phase winding within the superconductor's coherence length $\xi_{\rm sc}$. In comparison, the topological phase transition occurs for a phase gradient of roughly one phase winding per coherence length of the wire  $\xi_{\rm wire} = u/\Delta$. Thus, our scheme requires $\xi_{\rm sc} \lesssim \xi_{\rm wire}$.\cite{footnote} Both parameters entering the coherence length $\xi_{\rm sc} = v_F/\Delta_{\rm sc}$ of a clean superconductor, namely the Fermi velocity $v_F$ and the pairing amplitude $\Delta_{\rm sc}$, typically exceed the corresponding quantities for the wire. Thus, if both the wire and superconductor are clean, it is in general possible to satisfy the requirement by a suitable choice of materials and device parameters.  The estimates become yet more favorable for a dirty $s$-wave superconductor with coherence length $\xi=(\xi_{\rm sc}\ell)^{1/2}$, where $\ell$ is the elastic mean free path. 

We thank P. Brouwer, L. Jiang, and A.\ Morpurgo for useful discussions, and the Aspen Center for Physics for hospitality. We acknowledge support from the Deutsche Forschungsgemeinschaft through Priority Program 1285, the Moore Foundation and the National Science Foundation through the IQIM, as well as the National Science Foundation through grant DMR-1055522.


\begin{thebibliography}{12}
\bibitem{footnote} Ideally, this inequality should be strong so that the benefit of the smaller required Zeeman field is not compensated by a reduction in the superconductor's critical field. This also implies $ (\nabla \phi)^* \xi_{\rm{sc}} \ll 1$, so that the reduction of the magnitude $\Delta$ of the proximity induced gap due to the applied supercurrent remains negligible.
\end{thebibliography}
\end{document}